\documentclass[aps,pra,reprint,superscriptaddress]{revtex4-1}
\usepackage{amsfonts,amssymb,bm}
\usepackage{amsmath}
\usepackage{graphicx}
\usepackage{epsfig}
\usepackage{natbib}
\usepackage{mathrsfs}
\usepackage{color}
\usepackage{layout}
\begin{document}

\newcommand{\ket}[1]{\left| #1 \right\rangle}
\newcommand{\bra}[1]{\left\langle #1 \right|}
\newcommand{\braket}[2]{\left\langle #1 | #2 \right\rangle}
\newcommand{\braopket}[3]{\bra{#1}#2\ket{#3}}
\newcommand{\proj}[1]{| #1\rangle\!\langle #1 |}
\newcommand{\expect}[1]{\left\langle#1\right\rangle}
\newcommand{\Tr}{\mathrm{Tr}}
\def\Id{1\!\mathrm{l}}
\newcommand{\cM}{\mathcal{M}}
\newcommand{\cR}{\mathcal{R}}
\newcommand{\cE}{\mathcal{E}}
\newcommand{\cL}{\mathcal{L}}
\newcommand{\cl}{l}
\newcommand{\cH}{\mathcal{H}}
\newcommand{\cU}{\mathcal{U}}
\newcommand{\cP}{\mathcal{P}}
\newcommand{\reals}{\mathbb{R}}
\newcommand{\grad}{\nabla}
\newcommand{\rhohat}{\hat{\rho}}
\newcommand{\rhoMLE}{\rhohat_\mathrm{MLE}}
\newcommand{\rhotomo}{\rhohat_\mathrm{tomo}}
% The following line represents vectors as boldface.  Another possibility is via \vec{}.
\newcommand{\diff}{\mathrm{d}\!}
\newcommand{\pdiff}[2]{\frac{\partial #1}{\partial #2}}
\newcommand{\todo}[1]{\color{red}#1}
\def\FCW{1.0\columnwidth}
\def\HCW{0.55\columnwidth}
\def\TPW{0.33\textwidth}

%\graphicspath{{../Figures/}}



%\affiliation{University of Toronto}

%\thanks{}


%

%\pacs{42.50.Dv,42.50.Xa}
%\keywords{}


%\subsection{Overview}

\title{Quantum control of population transfer between vibrational states in an optical lattice}
\author{Matin Hallaji}
\email{mhallaji@physics.utoronto.ca}
\affiliation{Centre for Quantum Information and Quantum Control and Institute for Optical Sciences, Department of Physics, University of Toronto, Toronto, ON, M5S 1A7, Canada}
\author{Chao Zhuang}
\affiliation{Centre for Quantum Information and Quantum Control and Institute for Optical Sciences, Department of Physics, University of Toronto, Toronto, ON, M5S 1A7, Canada}
\author{Alex Hayat}
\affiliation{Department of Electrical Engineering, Technion, Haifa 32000, Israel}
\author{Felix Motzoi}
\affiliation{Department of Chemistry, University of California, Berkeley, CA 94720 USA}
\affiliation{Theoretical Physics, Saarland University, Campus, 66123 Saarbr\"ucken, Germany}
\author{Botan Khani}
\affiliation{IQC and Department of Physics and Astronomy, University of Waterloo, ON, N2L 3G1, Canada}
\author{Frank K. Wilhelm}
\affiliation{IQC and Department of Physics and Astronomy, University of Waterloo, ON, N2L 3G1, Canada}
\affiliation{Theoretical Physics, Saarland University, Campus, 66123 Saarbr\"ucken, Germany}
\author{Aephraim M. Steinberg}
\affiliation{Centre for Quantum Information and Quantum Control and Institute for Optical Sciences, Department of Physics, University of Toronto, Toronto, ON, M5S 1A7, Canada}
\date{\today}

\begin{abstract}
We study quantum control techniques, specifically Adiabatic Rapid Passage (ARP) and Gradient Ascent Pulse Engineering (GRAPE), for transferring atoms trapped in an optical lattice between different vibrational states. 
We compare them with each other and with previously studied coupling schemes in terms of performance. 
In our study of ARP, we realize control of the vibrational states by tuning the frequency of a spatial modulation through the inhomogeneously broadened vibrational absorption spectrum. We show that due to the presence of multiple crossings, the population transfer depends on the direction of the frequency sweep, in contrast to traditional ARP. In a second study, we control these states by applying a pulse sequence involving both the displacement of the optical lattice and modulation of the lattice depth.  This pulse is engineered via the GRAPE algorithm to maximize the number of atoms transferred from the initial (ground) state to the first excited state. We find that the ARP and the GRAPE techniques are superior to the previously tested techniques at transferring population into the first excited state from the ground state: $38.9\pm0.2\%$ and $39\pm2\%$ respectively. GRAPE outperforms ARP in leaving the higher excited states unpopulated (less than $3.3\%$ of the ground state population, at $84\%$ confidence level), while $18.7\pm0.3\%$ of the ground state population is transferred into higher excited states by using ARP. On the other hand, ARP creates a normalized population inversion of $0.21\pm0.02$, which is the highest obtained by any of the control techniques we have investigated.
\end{abstract}

\maketitle

\section{Introduction}

Control of coherent quantum states is a topic of ever-growing interest and plays an important role in various fields, such as atomic physics\cite{PhysRevLett.69.2353}, chemistry\cite{PhysRevLett.74.4799,PhysRevLett.92.113002}, biochemistry\cite{Kuroda09102009}, and condensed matter physics\cite{PhysRevLett.74.3596,PhysRevLett.78.306,DongSun2010,PhysRevB.61.7669,PhysRevB.69.195308,PhysRevLett.110.135505}. An attractive system for studying the control of quantum systems is an optical lattice, due to the ability to vary its system parameters in real time. An optical lattice is a periodic potential for atoms, formed by the interference pattern of laser beams\cite{OpticalLattice}. Since the lattice configuration and potential depth depend on the relative phase, polarization, and intensity of the laser beams, the Hamiltonian for atoms trapped in an optical lattice can be conveniently varied by adjusting laser parameters. In our experiment, we use the vibrational states of cold atoms trapped in a one-dimensional vertical optical lattice as a prototype system to study different quantum control techniques.

There have been several studies of the center-of-mass motion of atoms trapped in an optical lattice in different parameter regimes. Breathing-mode oscillations \cite{PhysRevLett.78.2928} were observed in deep lattice potentials ($U > 300E_{r}$, where $E_{r}$ is the effective recoil energy given by $E_{r} = \hbar^{2}k^{2}/(2m) = \hbar\omega_{r}$ with $k$ being the effective lattice wave vector and $m$ the mass of the atom) by suddenly increasing the lattice depth or by parametrically driving the lattice depth. Rabi oscillations between vibrational states \cite{PhysRevA.58.R2648} were observed in a very shallow lattice potential ($U \approx 6E_{r}$) by displacing the lattice potential sinusoidally. A demonstration of coupling vibrational states by combining sudden displacements of the lattice and delays between the displacements was presented in \cite{PhysRevA.77.022303}, and achieved the largest coupling coefficient between the lowest two vibrational states reported to date. It was demonstrated in \cite{TwoPath} that the combination of parametric driving of the lattice depth and sinusoidal displacement of the lattice potential results in quantum interference between different transition pathways, which was used to control the population transfer into different vibration states and suppress loss and leakage errors.

In this paper, we present two novel approaches for controlling vibrational states in an optical lattice and compare them with previous works. The first technique utilizes a frequency-chirped sinusoidal displacement of the lattice potential to couple the vibrational states, based on the concept of Adiabatic Rapid Passage (ARP) \cite{Powles1958,Camparo1984}. We experimentally study the dependence of the population transferred from the ground state into the first excited states on the sweep rate, the modulation amplitude and the sweep direction. The second technique involves coupling the vibrational states with both lattice displacement and modulation of the lattice depth, relying on the Gradient Ascent Pulse Engineering (GRAPE) algorithm \cite{Khaneja2005296} to optimize the coupling coefficient between the ground and the first excited state. Finally, we compare these two techniques to quantum control techniques previously applied to this system. Optimal population transfer and elimination of leakage error are of concern in many quantum information devices\cite{1367-2630-9-10-384} such as trapped ions\cite{PhysRevA.75.042329,benhelm2008towards}, cavity QED\cite{PhysRevA.72.032333}, superconductor qubits\cite{PhysRevA.82.040305,PhysRevLett.103.110501,PhysRevA.81.012306} and optical lattices\cite{anderlini2007controlled,PhysRevA.77.052309}. The results presented in this paper can be extended to those systems.

This paper is organized as follows. In Sec. \ref{secTheory} we describe the lattice Hamiltonian and the parameters available for controlling the quantum system. The experimental setup and the characterization of the system are shown in Sec. \ref{secStepup}. In Sec. \ref{secEarlyWork} we briefly revisit the techniques and results from our previous experiments. The experiment with the ARP technique is presented in Sec. \ref{secARP}. The experiment with the GRAPE technique is presented in Sec. \ref{secGRAPE}. In Sec. \ref{secComparison} we compare the performance of all quantum control techniques we have investigated, based on different figures of merit.

\section{Background Theory}\label{secTheory}

The Hamiltonian for an atom trapped in a one-dimensional optical lattice can be written as
\begin{equation}
  \mathbf{H_{0}} = \frac{\mathbf{p}^{2}}{2m} + U\sin^{2}k\mathbf{x},
  \label{eqnLatticeHamiltonian}
\end{equation}
where $m$ is the mass of the atom, $U$ is the lattice potential depth, and $k = \pi/d$ is the effective lattice wave vector, with $d$ being the lattice constant. The eigenstates of the Hamiltonian shown in Eq.(\ref{eqnLatticeHamiltonian}) are the well-known Bloch states\cite{kittel2005introduction}.

Atoms in our optical lattice can be treated as localized atoms in a single potential well, making it more convenient to decompose the state of the atom in the Wannier states \cite{PhysRev.115.809} basis than in the Bloch states basis. For the lattice depth we use, to good approximation, the lowest two bands are flat and the corresponding Wannier states are long lived. Therefore, they can be used as basis states that approximate energy eigenstates. By omitting the well index, we can write:
\begin{equation}
  \mathbf{H_{0}}|n\rangle \simeq E_{n}|n\rangle,
  \label{eqnWannierAsEigen}
\end{equation}
where $|n\rangle$ is the Wannier state for the n-th band, and $E_{n}$ is the average energy of the n-th band with $n=0$ being the ground state band. We label the resonance frequency of transition between the lowest two levels as $\omega_{01} = (E_{1} - E_{0})/\hbar$.

There are two operations that we use to couple the vibrational states in the optical lattice. The first is the displacement of the lattice; the time-dependent Hamiltonian for this operation is given by
\begin{equation}
\mathbf{H}(t) = \frac{\mathbf{p}^{2}}{2m} + U\sin^{2}\left[k\mathbf{x} + k\theta(t)\right],
    \label{eqnLabFrameDisplacementHamiltonian}
\end{equation}
where $\theta(t)$ is the function by which the lattice is displaced. We call this operation Phase Modulation (PM) hereafter. The other is the modulation of the lattice depth. The time-dependent Hamiltonian with lattice depth modulated is
\begin{equation}
\mathbf{H}(t) = \frac{\mathbf{p}^{2}}{2m} + \left[1+\eta(t)\right]U\sin^{2}k\mathbf{x},
    \label{eqnLabFrameDepthChangeHamiltonian}
\end{equation}
where $\eta(t)$ is the ratio of the change in lattice depth to the original lattice depth. We call this operation Amplitude Modulation (AM) hereafter. When both operations are applied together, the time-dependent Hamiltonian becomes
\begin{equation}
\mathbf{H}(t) = \frac{\mathbf{p}^{2}}{2m} + \left[1+\eta(t)\right]U\sin^{2}\left[k\mathbf{x} + k\theta(t)\right].
    \label{eqnLabFrameTimeDependentHamiltonian}
\end{equation}

The effect of this time-dependent Hamiltonian on the vibrational states is most easily understood in the reference frame of the moving optical lattice\cite{TwoPath}. In that reference frame, the time-dependent Hamiltonian is
\begin{equation}
  \mathbf{H_{lattice}}(t) = \frac{\mathbf{p}^{2}}{2m} + \left[1+\eta(t)\right]U\sin^{2}k\mathbf{x} - m\ddot{\theta}(t)\mathbf{x},
    \label{eqnLatticeFrameDisplacementHamiltonian}
\end{equation}
or simply
\begin{equation}
  \mathbf{H_{lattice}}(t) = \mathbf{H_{0}} + \eta(t)U\sin^{2}k\mathbf{x} - m\ddot{\theta}(t)\mathbf{x}.
    \label{eqnLatticeFrameDisplacementHamiltonianS}
\end{equation}
The connection between Eq. (\ref{eqnLabFrameTimeDependentHamiltonian}) and Eq. (\ref{eqnLatticeFrameDisplacementHamiltonianS}) can be easily understood from a classical point of view. In the reference frame of the displaced lattice, an atom sees a stationary lattice plus a fictitious gravitational force proportional to the acceleration of the moving lattice. Hence, the time-dependent Hamiltonian in the displaced lattice reference frame is simply the Hamiltonian of the stationary lattice shown in Eq. (\ref{eqnLatticeHamiltonian}) plus the change in the lattice depth and the fictitious potential, the second and the third terms in Eq. (\ref{eqnLatticeFrameDisplacementHamiltonianS}), respectively. A rigorous quantum version of the derivation reveals that the Hamiltonians in Eq. (\ref{eqnLabFrameTimeDependentHamiltonian}) and Eq. (\ref{eqnLatticeFrameDisplacementHamiltonianS}) lead to the same propagator (up to a global phase) from initial time $t_{i}$ to final time $t_{f}$, provided that $\theta(t_{i}) = \theta(t_{f}) = 0$ and $\dot{\theta}(t_{i}) = \dot{\theta}(t_{f}) = 0$ \cite{chao_thesis}. Eq. (\ref{eqnLatticeFrameDisplacementHamiltonianS}) shows that the coupling between the vibrational states by PM is mathematically equivalent to the dipole coupling of an electronic transition by a time-dependent electric field. In particular, this system is subject to a similar selection rule forbidding coupling between states of the same parity. The PM term, on the other hand, is incapable of coupling states of opposite parity.
%One special case of coupling the vibrational states is when the lattice is only sinusoidally displaced with $\theta(t) = a_{PM}\left(1-\cos\omega t\right)$, where $\omega$ is the driving frequency and $a_{PM}$ is the driving amplitude. From Eq. (\ref{eqnLatticeFrameDisplacementHamiltonianS}), the Hamiltonian in the displaced lattice reference frame becomes
%\begin{equation}
%  \mathbf{H}(t) = \mathbf{H_{0}} - ma_{PM}\omega^{2}\cos\omega t\mathbf{x},
%  \label{eqnMovingFrameTimeDependentHamiltonian}
%\end{equation}
%which means Rabi oscillation are expected between states $|n\rangle$ and $|n^{\prime}\rangle$, with the Rabi frequecny
%\begin{equation}
%  \Omega_{nn^{\prime}} = ma_{PM}\omega^{2}\langle n|\mathbf{x}|n^{\prime}\rangle/\hbar,
%  \label{eqnRabiFrequecny}
%\end{equation}
%when the approximation in Eq. (\ref{eqnWannierAsEigen}) is assumed.

\section{Experimental Setup}\label{secStepup}

We perform our experiment using cold $^{85}Rb$ atoms trapped in a one-dimensional optical lattice. This lattice is formed by two laser beams from the same laser source, intersecting at an angle of about $\theta = 50^{\circ}$ with parallel polarization to form a vertical interference pattern \cite{OpticalLattice}. The frequency of the laser is blue-detuned by 30GHz from the D2 line of $^{85}$Rb ($\lambda = 780.24$nm). At this frequency and incident angle, the lattice constant $d = \frac{\lambda}{2}/\sin\frac{\theta}{2}$ is roughly $1\mu$m. The recoil frequency ($\omega_r = E_{r}/\hbar$) can be calculated to be $\omega_{r} = 2\pi\times h/(8md^{2}) = 2\pi \times 685$Hz. The lifetime of the atoms trapped in our optical lattice is set by the photon scattering rate, estimated to be about $50$ms. This time is much longer than our experiment duration, which is typically around $1$ms. The sample of $^{85}Rb$ atoms used in these experiments is laser-cooled to approximately $10\mu$K, with a density sufficiently low ($<10^{10}$ atoms/cm$^3$) that interactions between atoms can be neglected. At  these temperatures the thermal de Broglie wavelength of the atoms is around $60$nm, which is much shorter than the lattice constant. Hence there is no coherence between atoms trapped in neighboring wells, and we are free to think of each trapped atom as being localized in one potential well.

Since our optical lattice is in the vertical direction, the potential trapped atoms feel is the sum of the periodical optical lattice potential and the linear gravitational potential. In our case, the gravitational potential is about $2.86E_{r}$ per lattice site. This adds an additional energy term $mg\mathbf{x} = 2.86E_r \mathbf{x}/d$ to the Hamiltonian in Eq. (\ref{eqnLatticeFrameDisplacementHamiltonianS}), where $g$ is the gravitational acceleration.

We make use of the gravitational potential for state preparation and measurement. By adiabatically lowering the depth of the optical lattice until only one vibrational state is supported, and then adiabatically increasing it again, we are able to prepare the atoms in the lowest vibrational state. This same filtering technique\cite{PhysRevA.72.013615} is used to measure the populations in different vibrational states after excitation. Our population measurement only discriminates the lowest two vibrational states from the higher ones. We refer to the normalized population in the ground state, the first excited state, and all other states as $P_{0}$, $P_{1}$, and $P_{L}$, respectively. Due to imperfections in the preparation stage, there is always an initial population in $P_{1}$. Experimentally, we are able to keep most of the population in the ground state, $4-8\%$ of total population in the first excited state, and nothing detectable in higher excited states. We label the initial population in the ground and the first excited state as $P_{0}^{i}$ and $P_{1}^{i}$, respectively.

To experimentally realize the coupling methods in Eq. (\ref{eqnLabFrameDisplacementHamiltonian}) and Eq. (\ref{eqnLabFrameDepthChangeHamiltonian}), we use acousto-optic modulators (AOMs) to phase modulate one and amplitude modulate the other laser beam that form the lattice. PM of one laser beam results in a displacement of the lattice potential and AM of the other laser beam results in a modulation of the lattice depth.

The laser beam we use has a Gaussian intensity distribution. Therefore, atoms located at different horizontal positions experience different lattice depths due to the different laser intensities. In our case, the measured Gaussian intensity profile of the laser beam has an r.m.s. radius around $1.5mm$. 
Although the lattice depth distribution can be estimated from the laser parameters and the shape of the atomic cloud, but precise characterization of it requires experimental measurement.

\begin{figure}[!ht]
  \includegraphics[width=\columnwidth]{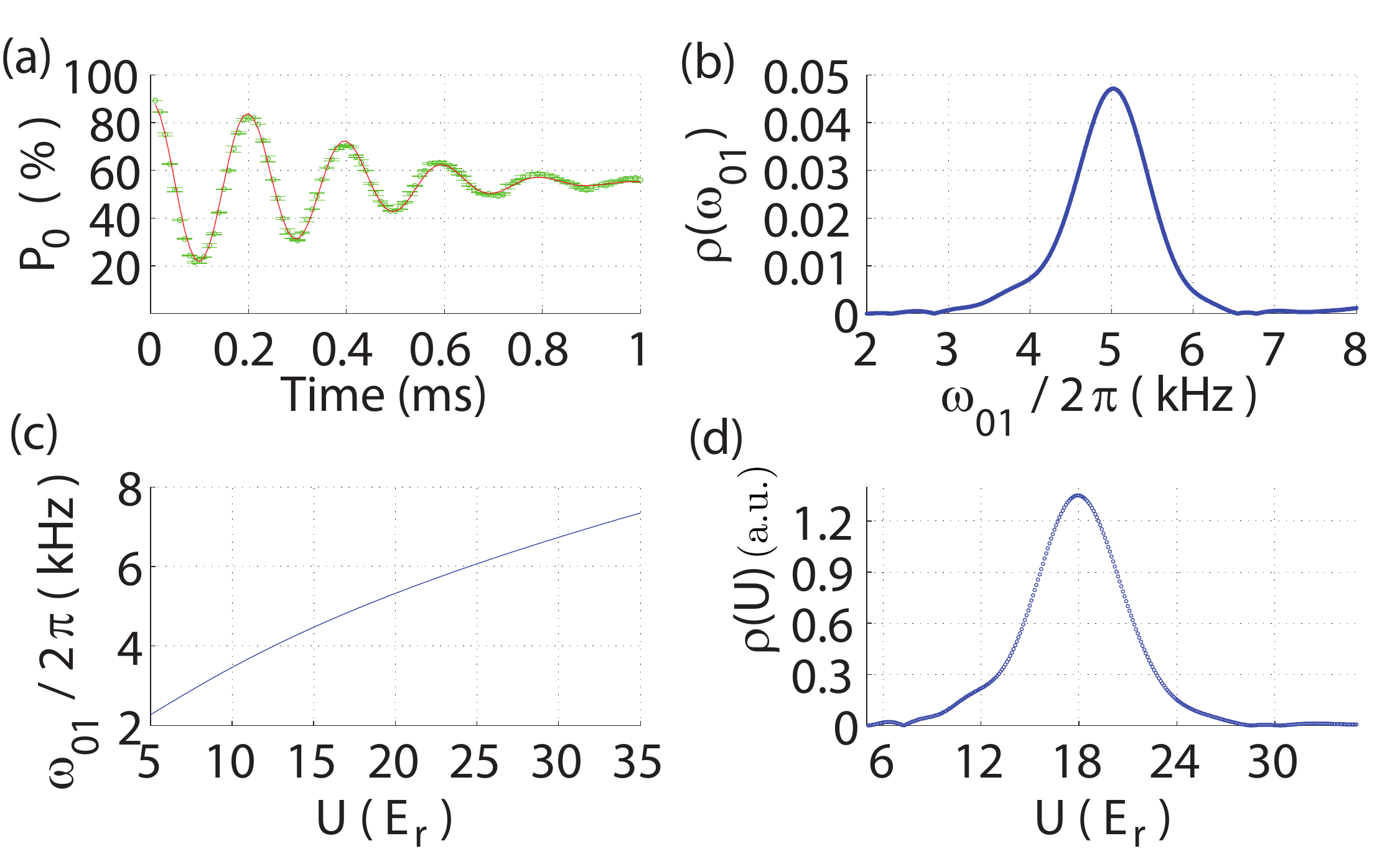}
  \caption{\label{figLatticeDepthDistribution} Experimentally measured lattice depth distribution. (a) Ground state population vs. time delay in a Ramsey spectroscopy experiment. (b) Probability density of atoms oscillating at different frequency $\omega_{01}$. (c) Relation between $\omega_{01}$ and lattice depth $U$. (d) Lattice depth distribution derived from (b) and (c).}
\end{figure}

In order to estimate the inhomogeneous broadening of the system experimentally, we perform a Ramsey-type interference experiment\cite{PhysRevA.77.022303}. To start, we prepare atoms in the ground state of the optical lattice. Then we abruptly displace the optical lattice (in less than $0.5\mu s$). After a variable time delay, a second abrupt displacement is made in the opposite direction to move the optical lattice back to its original configuration. Finally, we measure the ground state population versus the time delay, Fig. \ref{figLatticeDepthDistribution}(a).
The green dots are the experimentally measured results and the solid red line is the fitting curve. Evidently, the population coherently oscillates in and out of the ground state. The amplitude of this oscillation decays because of the inhomogeneous broadening. The formula used for the fitting is
\begin{equation}
  P_{g}^{s} = A_{s}e^{-(\gamma_{2}^{s}t)^{2}/2}\cos(\omega_{s}t+\phi_{s})+B_{s}.
\end{equation}
The fitting results show $\gamma_{2}^{s} = 2.91\pm0.03$KHz and $\omega_{s} = 5.007\pm0.008$KHz. The value of $\omega_{s}$ corresponds to a lattice depth of $18E_{r}$. In order to obtain the lattice depth distribution, first we find the frequency distribution of the measured oscillation by taking its Fourier transform, Fig. \ref{figLatticeDepthDistribution}(b). By knowing the relationship between lattice depths and resonant frequencies ($\omega_{01}$), as shown in Fig. \ref{figLatticeDepthDistribution}(c), one can find the find the lattice depth distribution to be $\rho(U) = \rho(\omega_{01})\frac{\mathrm{d}\omega_{01}}{\mathrm{d}U}$. Fig. \ref{figLatticeDepthDistribution}(d) shows the derived lattice depth distribution for our ARP experiment.

\section{Experiment Results and Analysis}

\subsection{Previous Experiments and Numerical Simulation}\label{secEarlyWork}

In this section, we briefly review the different previous experiments on coupling vibrational states in our experimental setup and the numerical method we use to simulate the experimental results.

In \cite{PhysRevA.77.022303}, we studied different combinations of displacements and time delays in order to optimize the transition probability from the ground state to the first excited state. In that work, three different kinds of pulses were studied: the single-step pulse (where $\theta(t)$ is a Heaviside step function times an amplitude $a_{PM}$), the square pulse (where $\theta(t)$ is a rectangular function with pulse duration of $t_{p}$ and amplitude of $a_{PM}$), and the Gaussian pulse (where $\theta(t)$ is a Gaussian function with full-width-at-half-maximum temporal width of $t_{p}$ and peak amplitude of $a_{PM}$). The best experimentally measured transition probability between the ground and the first excited state was found to be $P_{1} = 33\pm1\%$, with $P_{L} = 30\pm1\%$, via the Square pulse. If the vibrational states are assumed to be those of a simple harmonic oscillator, then the maximum $P_{1}$ would be $1/e$ for any excitation created by combinations of displacements and time delays. Although simulations showed that $P_{1}$ could be higher than $1/e$ in the optical lattice, the experimental results did not surpass that bound due to inhomogeneous broadening.

%Another experiment in our setup measures the Rabi oscillations between different vibrational states\cite{Rabi}. In that work, by sinusoidally displacing the lattice with $\theta(t) = a_{PM}\left(1-\cos\omega t\right)$, Rabi oscillation between the lowest two vibrational states was observed with both mean and amplitude decay as a function of time. It was found that the decay of the mean is due to the gravity. And the decay of the amplitude is due to the inhomogeneous broadening. Oscillation in $P_{L}$ was observed as well, which leads to a consideration of simulation with the 3-level approximation of the time-dependent Hamiltonian Eq. (\ref{eqnMovingFrameTimeDependentHamiltonian}):
%\begin{equation}
%\begin{aligned}
%\mathbf{H_{3lvl}}(t) = &\sum_{n=0}^{2}E_{n}|n\rangle\langle n| + \hbar\Omega_{01}\left(|0\rangle\langle 1|+ |1\rangle\langle 0|\right)\cos\omega t\\
%&+ \hbar\Omega_{12}\left(|1\rangle\langle 2| + |2\rangle\langle 1|\right) \cos\omega t.
%\end{aligned}
% \label{eqnThreeLevelHamiltonian}
%\end{equation}
%The simulation using Eq. (\ref{eqnThreeLevelHamiltonian}) with consideration of the inhomogeneous broadening gives results which agree well with experimental results once the decay of mean is numerically removed.
A second approach that was previously investigated relied on interference between one-phonon and the two-phonon transitions to coherently control the coupling into different vibrational states\cite{TwoPath}. In that study, the goal was to optimize the branching ratio $P_1 / P_L$, that is to optimize the population transfer from the ground states to the first excites state while minimizing the loss. The two interfering paths from the ground state $P_0$ to the loss state $P_L$ were created by a sinusoidal displacement (PM) of the lattice, which couples $P_0$ to $P_L$ through $P_1$ (two-phonon transition), and a sinusoidal modulation of lattice depth (AM), which directly couples $P_0$ to $P_L$ (one-phonon transition.) The relative phase between the two paths was controlled by controlling the relative phase between AM and PM. Using this technique, we achieved a branching ratio of $17\pm2$, the highest achieved among quantum control techniques that used analogous schemes\cite{PhysRevB.39.3435}.

In this paper, to model the experiments, numerical simulations are conducted and the results are compared to the experimental data. A split-operator method is utilized to numerically solve the time-dependent Schr\"{o}dinger equation with the Hamiltonian shown in Eq. (\ref{eqnLabFrameTimeDependentHamiltonian}) for a certain lattice depth $U$. To take the initial population distribution into account, the time-dependent Schr\"{o}dinger equation is solved twice with initial state set to be $|0\rangle$ and then $|1\rangle$. The solutions are then averaged with weights of $P_{0}^{i}$ and $P_{1}^{i}$. When gravity is considered, the time-dependent Schr\"{o}dinger equation is solved with the term $mg\mathbf{x}$ added to the Hamiltonian. To include the effect of inhomogeneous broadening, the solutions of a range of different lattice depths are averaged according to the experimentally measured lattice depth distribution, e.g., Fig. \ref{figLatticeDepthDistribution}(d). For the previous experiments and experiments presented in this paper, simulation results with different conditions as mentioned above are compared to the experimental results.

\subsection{Adiabatic Rapid Passage}\label{secARP}

In this section, we present the results of an attempt to use Adiabatic Rapid Passage (ARP), well known to be robust against inhomogeneous broadening\cite{Powles1958,Camparo1984}, to improve the efficiency of coupling vibrational states in an optical lattice. We observe interesting features due to the multi-level nature of this system.

ARP for a 2-level system has been studied theoretically and experimentally in different systems, such as magnetic resonance\cite{PhysRev.70.460,PhysRev.70.474}, ion traps\cite{PhysRevA.58.2242,PhysRevA.84.033412,RAP100}, and Rydberg atoms\cite{PhysRevLett.96.073002}. By applying an ARP pulse in a 2-level system, transfer efficiency from the ground state to the excited state very close to $100\%$ can be achieved\cite{RAP100}. However, ARP in a multi-level system has additional features that are absent in a 2-level system\cite{PhysRevA.72.053404,PhysRevA.77.023406}.

\begin{figure}[!ht]
  \centering
  \includegraphics[width=\columnwidth]{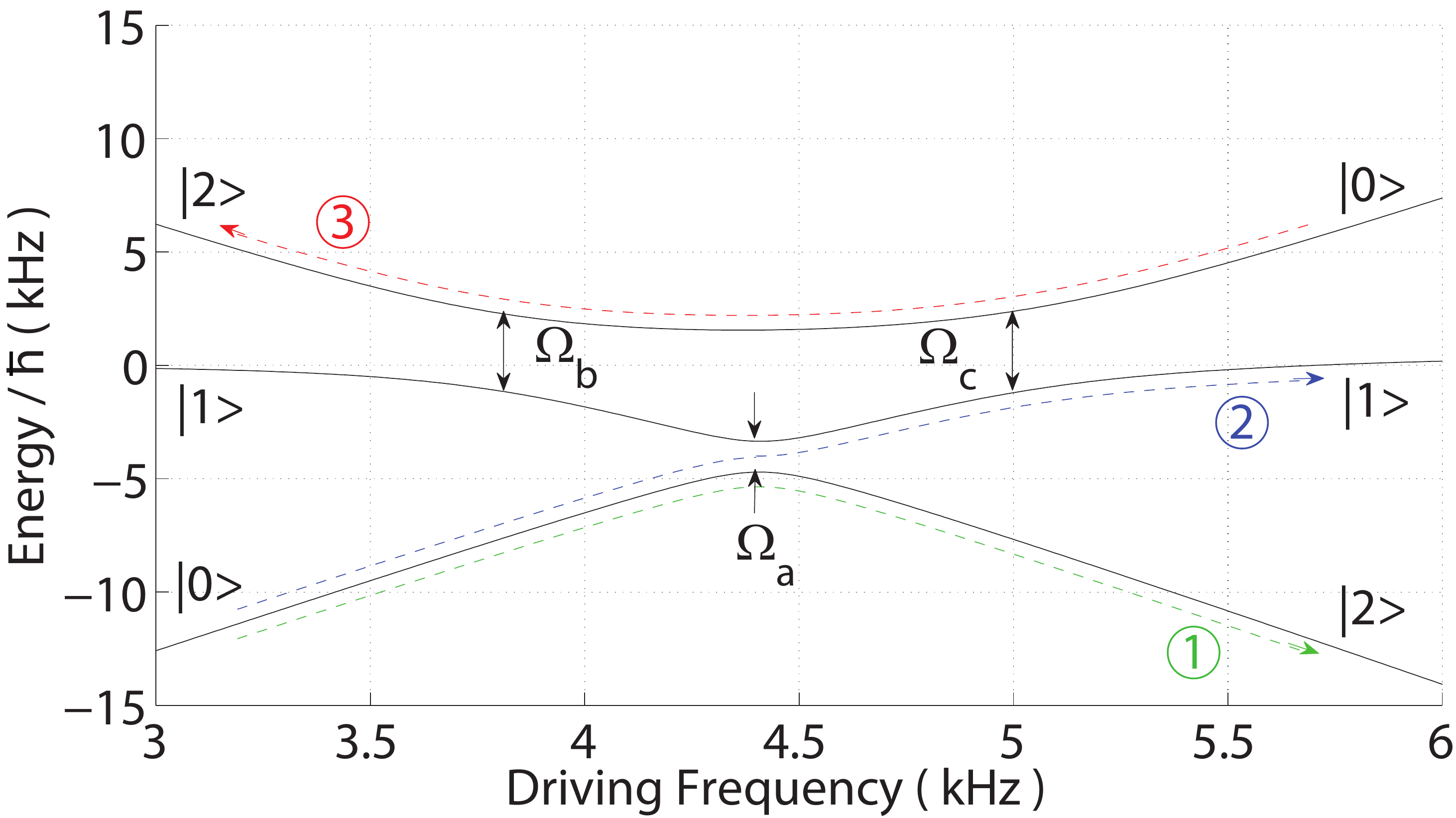}
  \caption{\label{fig3levelDressedLevels} Dressed state picture for a 3-level system. The dashed curves indicate different paths along which the state of the system evovles when ARP is applied. Path 1 and 3 (green and red respectively) are the adiabatic paths in which the system goes from ground to the second excited state and vice versa. Path 2, shown in blue, is when the chirp rate is adiabatic with resoect to $\Omega_b$ and $\Omega_c$ but not with respect to $\Omega_a$ and the system goes from the ground state to first excited for an up chirp.}
\end{figure}

To simplify the study of ARP in our system, we approximate the Hamiltonian in Eq.\ref{eqnLabFrameDisplacementHamiltonian} (with $\theta(t) = a_{PM}[1-\cos(\omega t )]$) by a three-level system Hamiltonian given by
\begin{equation}
\begin{aligned}
\mathbf{H_{3lvl}}(t) = &\sum_{n=0}^{2}E_{n}|n\rangle\langle n| + \hbar\Omega_{01}\left(|0\rangle\langle 1|+ |1\rangle\langle 0|\right)\cos\omega t\\
&+ \hbar\Omega_{12}\left(|1\rangle\langle 2| + |2\rangle\langle 1|\right) \cos\omega t.
\end{aligned}
 \label{eqnThreeLevelHamiltonian}
\end{equation}
with $|0\rangle$, $|1\rangle$ and $|2\rangle$ being the ground state, first excited state and second excited state respectively. Fig. \ref{fig3levelDressedLevels} shows the dressed state picture for the 3-level system approximating the lowest three levels of an optical lattice, with lattice depth of $18E_{r}$ and a driving amplitude of $a_{PM} = d/36$, versus the driving frequency.
There are three energy gaps (avoided crossings) in Fig. \ref{fig3levelDressedLevels}, which are labeled with the corresponding Rabi frequency $\Omega_{a}$, $\Omega_{b}$, and $\Omega_{c}$. The gap $\Omega_{a}$ is due to the second order coupling between $|0\rangle \rightarrow |2\rangle$, that results in a much smaller gap than the others (labeled $\Omega_{b}$ and $\Omega_{c}$), which are caused by direct coupling between $|1\rangle \rightarrow |2\rangle$ and $|0\rangle \rightarrow |1\rangle$ respectively.

To perform an ARP pulse, the system is driven with a frequency far detuned from $\omega_{01}$. Then the driving frequency is swept through $\omega_{01}$ to end far detuned on the other side. The chirp from red detuning to blue detuning is referred to as an up chirp, and the chirp from blue detuning to red detuning is referred to as a down chirp. There are many routes that the system could take depending on the chirp rate $|\beta|$ and the chirp direction (up or down.) Among all these routes, the most interesting three routes are labeled and shown by dashed curves in Fig. \ref{fig3levelDressedLevels}. Route \textcircled{1} and route \textcircled{3} show that if the initial state is $|0\rangle$ and the chirp is adiabatic with respect to all the crossings ($|\beta| \ll \Omega_{a}^2 , \Omega_{b}^2 , \Omega_{c}^2$), the final state will be $|2\rangle$, independent of the chirp direction. On the other hand, route \textcircled{2} shows that if the chirp is adiabatic only with respect to the crossings $\Omega_{b}$ and $\Omega_{c}$ ($|\beta| \ll \Omega_{b}^2$ and $|\beta| \ll \Omega_{c}^2$), but non-adiabatic with respective to the crossing $\Omega_{a}$ ($|\beta| \gg \Omega_{a}^2$), for initial state of $|0\rangle$ the final state is going to be $|1\rangle$ for an up chirp, whereas for the down chirp the final state will be $|2\rangle$. This direction dependence is essentially different from the ARP for a 2-level system and provides us with the sweep direction as a control parameter, besides $|\beta|$, to control the coupling the ground state into different final states.

To realize this ARP technique experimentally in the optical lattice, PM with $\theta(t) = a_{PM}\left[1-\cos\left(\omega_{i} t + \frac{1}{2}\beta t^{2} \right)\right]$ is applied to perform a frequency chirped pulse, where $\omega_{i}$ is the initial driving frequency, and $|\beta|$ is the chirp rate (positive $\beta$ corresponding to up chirp and negative $\beta$ corresponding to down chirp). We perform the experiment for all combinations of pulse durations $t_{p} = [0.4; 0.6; 0.8; 1; 2; 3; 4; 5]$ms and covered frequency ranges $\Delta_{f} = [3.6; 4.8; 6; 7.2; 9]$kHz. For each pair of pulse duration and frequency coverage, the chirp rate is given by $|\beta| = 2\pi\Delta_f / t_p$. The center frequency of the chirped pulse $\omega_{c}$ is set to be $2\pi\times5kHz$, which is the center resonance frequency of the lattice, as shown in Fig. \ref{figLatticeDepthDistribution}(d).  

%The relationship between the frequency range covered, $\Delta_{f}$, and the pulse duration, $t_{p}$, is $2\pi\Delta_{f} = |\beta| t_{p}$. By varying $\Delta_{f}$ and $t_{p}$, the chirp rate $|\beta|$ can be varied. To keep the experiment simple, $t_{p}$ is chosen such that $\omega_{i} t_{p} + \frac{1}{2}|\beta| t_{p}^{2} = 2j\pi$, where $j$ is an integer. As the measured lattice depth distribution for this part of the experiment is shown in Fig. \ref{figLatticeDepthDistribution}(d), the center of the frequency chirp $\omega_{c}$ is set to be $2\pi\times5kHz$. Experimentally, combinations of $\Delta_{f} = 3.6; 4.8; 6; 7.2; 9kHz$ and $t_{p} = 0.4; 0.6; 0.8; 1; 2; 3; 4; 5ms$ for both up and down chirps are measured.

\begin{figure*}[!ht]
  \centering
  \includegraphics[width=\textwidth]{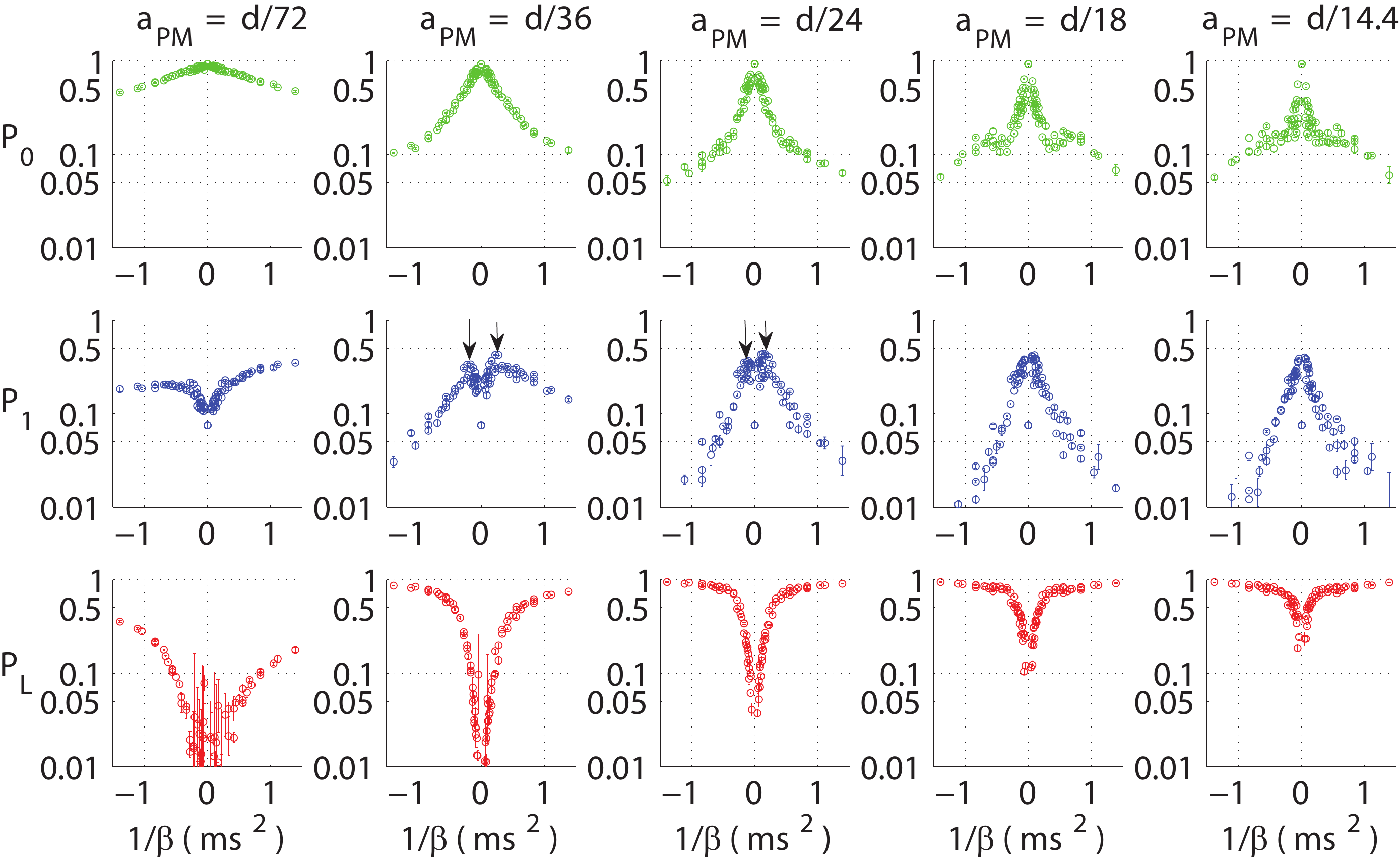}
  \caption{\label{figARPresults} Experimentally measured population transfers vs chirp rates. The green, blue and red are ground state, excited state and leaked states respectively. The driving amplitude increases from left to right. The arrows point to the maximum population transfer to the first excited state from the ground state, for displacement amplitudes of d/36 and d/24, where d is the lattice constant.}
\end{figure*}

In order to examine the connection between the measured ARP results and the Landau-Zener (LZ) non-adiabatic transition\cite{Camparo1984,Zener1932,PhysRevA.77.023406}, the measured population is plotted against $1/\beta$ on a semi-log plot in Fig. \ref{figARPresults}: $P_{0}$, $P_{1}$, and $P_{L}$ are shown in the top, middle and bottom rows, respectively; each column shows the results of different driving amplitudes, increasing from left to right.
Since $1/|\beta| = \frac{t_{p}}{2\pi\Delta_{f}}$, the point at $1/\beta = 0$ corresponds to either $t_{p} = 0$ or $\Delta_{f} \rightarrow \infty$. The former case, corresponding to a zero-duration pulse, obviously leads to no population transfer; it turns out that since the latter case is extremely non-adiabatic ($|\beta| \gg \Omega_{a}^2$,$\Omega_{b}^2$,$\Omega_{c}^2$), it does not excite the system either. Thus the points at $1/\beta = 0$ in Fig. \ref{figARPresults} correspond to the measured initial populations of $P_{0}^{i} = 92.5\pm0.5\%$ and $P_{1}^{i} = 7.5\pm0.5\%$. According to LZ formula\cite{Zener1932}, the probability of transferring population from one state to another is 
%The former case does not excite in the system and the later case gives an extremely non-adiabatic chirp ($|\beta| \gg \Omega_{a}^2$, $|\beta| \gg \Omega_{b}^2$, and $|\beta| \gg \Omega_{c}^2$) leading to no excitation in the system as well, which means the points at $1/\beta = 0$ in Fig. \ref{figARPresults} correspond to the measured initial populations of $P_{0}^{i} = 92.5\pm0.5\%$ and $P_{1}^{i} = 7.5\pm0.5\%$. For a 2-level system, the LZ formula reads\cite{Zener1932},
\begin{equation}
  P = \exp{\left(-\frac{1}{4}\frac{\Omega^{2}}{|\beta|}\right)},
  \label{eqnLZ2level}
\end{equation}
where $\Omega$ is the Rabi frequency at the level crossing. Evidently, this probability is independent of the chirp direction. Fig. \ref{figARPresults}, however, shows dependence of population transfer on the chirp direction, which is the signature of presence of multiple crossings. This dependence is most obvious for the $P_{1}$ vs. $1/\beta$ curves (the middle row), where for the same chirp rate, the up chirp gives higher $P_{1}$ than the down chirp. For example, in the subplot for a driving amplitude of $a_{PM} = d/36$, the maximum $P_{1}$ for the up chirp is $42.5\pm0.4\%$, higher than the maximum $P_{1}$ for the down chirp, $33.7\pm0.6\%$ (both indicated by arrows in Fig. \ref{figARPresults}.) In the subplot for driving amplitude of $a_{PM} = d/24$, the maximum $P_{1}$ for the up chirp is $43.5\pm0.2\%$, higher than the maximum $P_{1}$ for the down chirp, $36.7\pm0.4\%$. In both cases this maximum exceeds the $1/e$ bound for coupling vibrational states with displacement operators in a simple harmonic trap. One can also easily see that for small driving amplitudes ($a_{PM} = d/72$, $d/36$, and $d/24$), the up chirp gives lower $P_{L}$ than the down chirp as shown in the bottom row of Fig. \ref{figARPresults}. This asymmetric behavior can be qualitatively explained with the 3-level ARP model in Fig. \ref{fig3levelDressedLevels}, where different chirp directions lead to different final states.

\begin{figure*}[!ht]
  \centering
  \includegraphics[width=\textwidth]{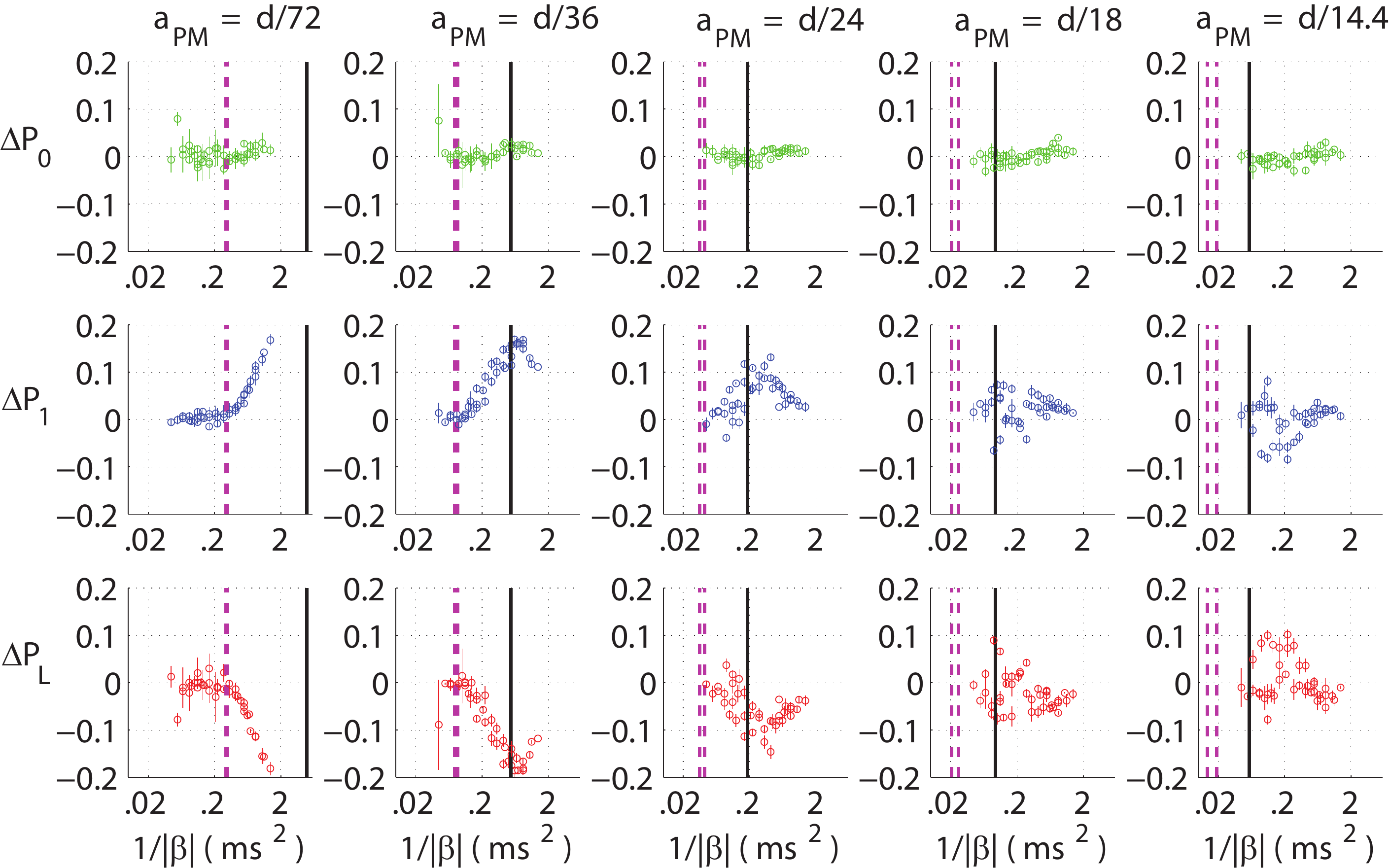}
  \caption{\label{figUpVsDownLogScale} Population differences between up chirp and down chirp vs magnitude of chirp rate. The purple dashed lines indicate $1 / \Omega_b ^2$ and $1 / \Omega_c ^2$ and the solid black line indicates $1 / \Omega_a ^2$ as defined in Fig. \ref{fig3levelDressedLevels}. Note that the asymmetry is only observed when $1/|\beta|$ is greater than $1 / \Omega_b ^2$ and $1 / \Omega_c ^2$ but not much greater than $1 / \Omega_a ^2$.}
\end{figure*}

To further compare the difference between the up chirp and the down chirp to the adiabatic criteria, the population difference between the up chirp and the down chirp is plotted vs. $1/|\beta|$ in Fig. \ref{figUpVsDownLogScale}.
The top, middle, and bottom rows in Fig. \ref{figUpVsDownLogScale} show $\Delta P_{0}$ (population difference for the ground state), $\Delta P_{1}$ (population difference for the first excited state), and $\Delta P_{L}$ (population difference for the leakage), respectively. Each column in Fig. \ref{figUpVsDownLogScale} shows the results of different driving amplitudes, increasing from left to right. The black solid line denotes $1/\Omega_{a}^{2}$ for the corresponding driving amplitude. To the left of this line are chirp rates that are relatively non-adiabatic with respect to the crossing $\Omega_{a}$ (i.e., $1/|\beta| < 1/\Omega_{a}^{2}$), as shown in Fig. \ref{fig3levelDressedLevels}. The two purple dashed lines denote $1/\Omega_{b}^{2}$ and $1/\Omega_{c}^{2}$ for the corresponding driving amplitude. To the right of these lines are chirp rates that are relatively adiabatic ($1/|\beta| > 1/\Omega_{b}^{2}$ and $1/|\beta| > 1/\Omega_{c}^{2}$) with respect to the crossings $\Omega_{b}$ and $\Omega_{c}$. The top row of Fig. \ref{figUpVsDownLogScale} shows that $\Delta P_{0}$ is always close to zero and does not depend on $|\beta|$, which means the up and down chirp ARP pulses excite the same amount of population out of the ground state for the same chirp rate. The middle ($\Delta P_{1}$) and the bottom ($\Delta P_{L}$) rows of Fig. \ref{figUpVsDownLogScale} both have three regimes. Where the chirp is relatively non-adiabatic with respect to all the crossings ($1/|\beta| < 1/\Omega_{b}^{2}$, $1/|\beta| < 1/\Omega_{c}^{2}$, and $1/|\beta| < 1/\Omega_{a}^{2}$), i.e., left of the two purple dashed lines in Fig. \ref{figUpVsDownLogScale}, $\Delta P_{1}$ ($\Delta P_{L}$) is zero and does not depend on $|\beta|$. As the chirp rate $|\beta|$ becomes relatively adiabatic with respect to the $\Omega_{b}$ and $\Omega_{c}$ crossings ($1/|\beta| > 1/\Omega_{b}^2$ and $1/|\beta| > 1/\Omega_{c}^{2}$) but remains non-adiabatic with respect to the $\Omega_{a}$ crossing ($1/|\beta| < 1/\Omega_{a}^{2}$), the difference between the up and down chirp becomes significant. As $|\beta|$ grows more adiabatic with respect to the $\Omega_{b}$ and $\Omega_{c}$ crossings (further away from the purple dashed lines to the right in Fig. \ref{figUpVsDownLogScale}), the difference between the up and down chirps becomes more significant: $\Delta P_{1}$ becomes more positive and $\Delta P_{L}$ becomes more negative as $\Omega_{b}^{2}/|\beta|$ ($\Omega_{c}^{2}/|\beta|$) gets larger. Once the chirp rate $|\beta|$ becomes adiabatic with respect to all the crossings ($1/|\beta| > 1/\Omega_{b}^{2}$, $1/|\beta| > 1/\Omega_{c}^{2}$, and $1/|\beta| > 1/\Omega_{a}^{2}$), i.e., to the right of the black solid line, the difference between the up and down chirp starts to diminish. As one can see $\Delta P_{1}$ becomes less positive and $\Delta P_{L}$ becomes less negative as $\Omega_{a}^{2}/|\beta|$ gets larger on the right side of the black solid line in Fig. \ref{figUpVsDownLogScale}. The measured dependence of $\Delta P_{1}$ ($\Delta P_{L}$) on the chirp rate $|\beta|$ agrees with our expectation from the 3-level ARP model: when the chirp is adiabatic with respect to the crossings $\Omega_{b}$ and $\Omega_{c}$ but non-adiabatic with respect to the crossing $\Omega_{a}$, up chirp transfers more population into the first excited state than the down chirp.

\begin{figure*}[!ht]
  \centering
  \includegraphics[width=\textwidth]{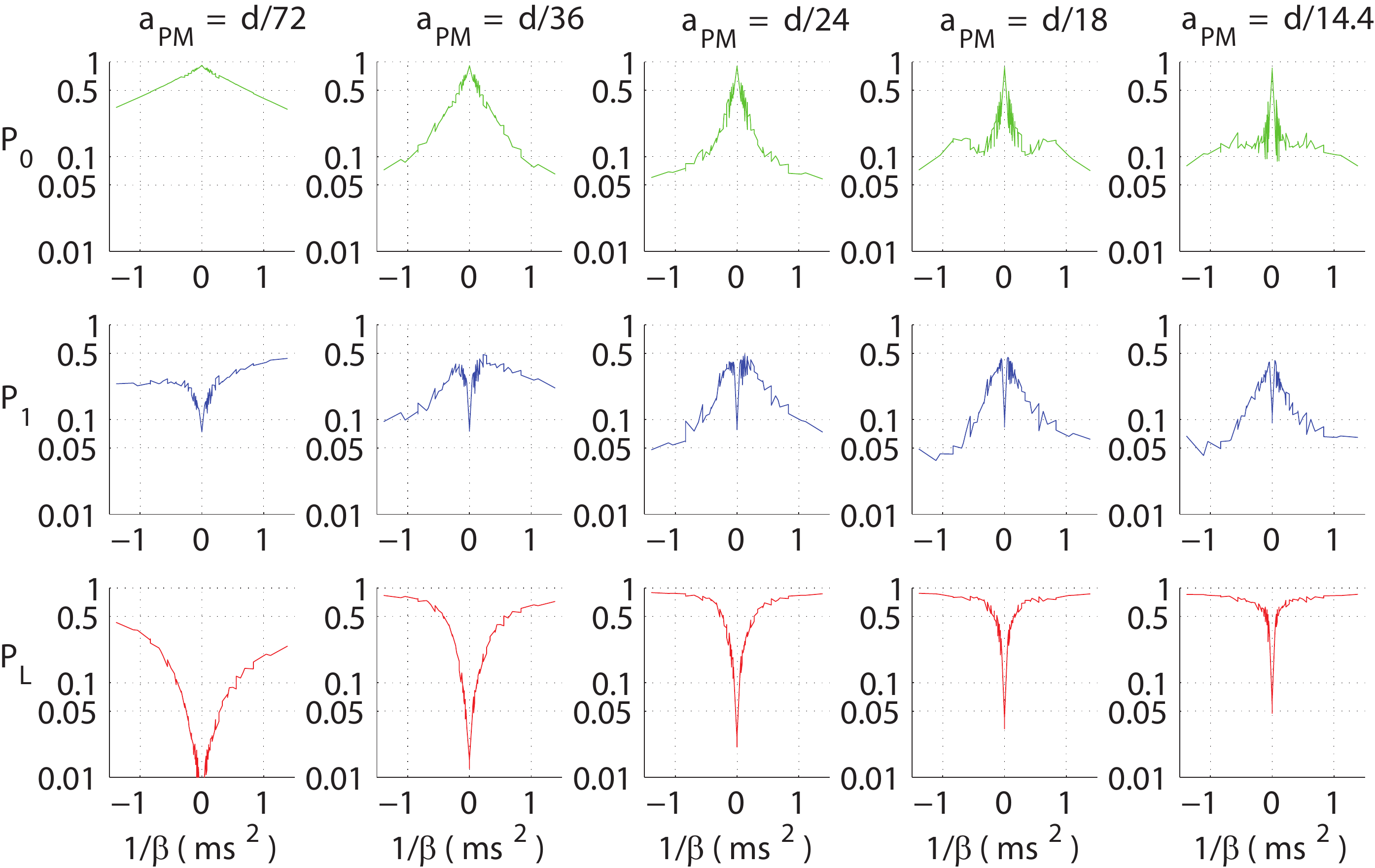}
  \caption{\label{figARPFullSimulation} Simulation results of ARP using the lattice Hamiltonian including the effect of gravity. The effects of inhomogeneous broadening and imperfect initial population are considered in these simulations as well.}
\end{figure*}

The other difference between the 3-level ARP and the 2-level ARP is in the relationship between the excited population and $1/|\beta|$ for a given chirp direction. With the Landau-Zener formula in Eq. (\ref{eqnLZ2level}), the 2-level ARP model predicts a straight line in the semi-log plot of populations vs. $1/|\beta|$. The measured results of the 3-level ARP in Fig. \ref{figARPresults} mostly show two lines and some transition stage between them (on either the up chirp side or the down chirp side), indicating that a Landau-Zener-type formula could still be useful to explain the results. In fact, a theoretical study\cite{PhysRevA.77.023406} has been conducted using ``analytic approximation by assuming independent pairwise Landau-Zener transitions occurring instantly at the relevant avoided crossings'' to explain the 3-level ARP. In that work, the authors showed an analytical solution which agrees well with numerical simulation results under the assumption of equal coupling between adjacent levels ($\Omega_{01} = \Omega_{12}$ and $\Omega_{02} = 0$). However, this assumption is not applicable to our system (as $\Omega_{01} \neq \Omega_{12}$), preventing the direct application of their analytical solution. But the similarity between their numerical simulation results and the experimental results here indicates that a similar analytic solution by assuming independent pairwise Landau-Zener transitions could potentially be used to explain the experimental results. Instead of trying to find the analytical solution for this system, a simulation with the full lattice Hamiltonian including the gravity and the inhomogeneous broadening is performed. The simulation results shown in Fig. \ref{figARPFullSimulation} agree well with the experimental data shown in Fig. \ref{figARPresults}.

\subsection{Gradient Ascent Pulse Engineering}\label{secGRAPE}

Since its invention in 2005, the GRadient Ascent Pulse Engineering (GRAPE)\cite{Khaneja2005296} technique has been applied to design optimal control pulse sequences in different quantum systems, such as: magnetic resonance\cite{souzaexperimental2011,PhysRevLett.107.170503,PhysRevLett.104.160501,PhysRevA.84.042312}, superconducting qubits\cite{PhysRevB.79.060507,PhysRevLett.103.110501}, circuit QED\cite{PhysRevB.85.054504}, and nitrogen-vacancy centers\cite{PhysRevA.80.032303}. In this section we discuss the application of GRAPE to control the vibrational states of atoms trapped in an optical lattice. The set of controls available in our experiment consists of displacements of the lattice and modulations of the lattice depth. We refer to the temporal profile of the displacement and modulation as a pulse sequence. The algorithm works by evaluating the fidelity function between a desired propagator and the propagator for a given pulse sequence. The analytically obtained gradient of the fidelity function is then used to update the pulse sequence by moving along the direction of the gradient. This is then repeated until the desired propagator is achieved.

To account for the lattice dispersion relation in our system, a new version of GRAPE has been developed \cite{PhysRevA.85.022306}. This version of GRAPE calculates the propagators for the transition between the ground Bloch state and the exited Bloch state for every quasi-momentum. The spread in these propagators is due to the different energy differences (diagonal elements of the transition matrix) which come from the difference in quasi-momenta and to the different control elements (off-diagonal elements of the transition matrix) which come from different coupling coefficients. The net fidelity and gradient of the ensemble is the average of all the propagators calculated. Taking the above procedure, the robust version of GRAPE generates a pulse sequence which efficiently couples every pair of Bloch states. By setting the $X_{\pi}$-gate of the lowest two vibrational states as the desired propagator, pulse sequences are generated with an averaged fidelity close to one.

\begin{figure*}[!ht]
  \centering
  \includegraphics[width=\textwidth]{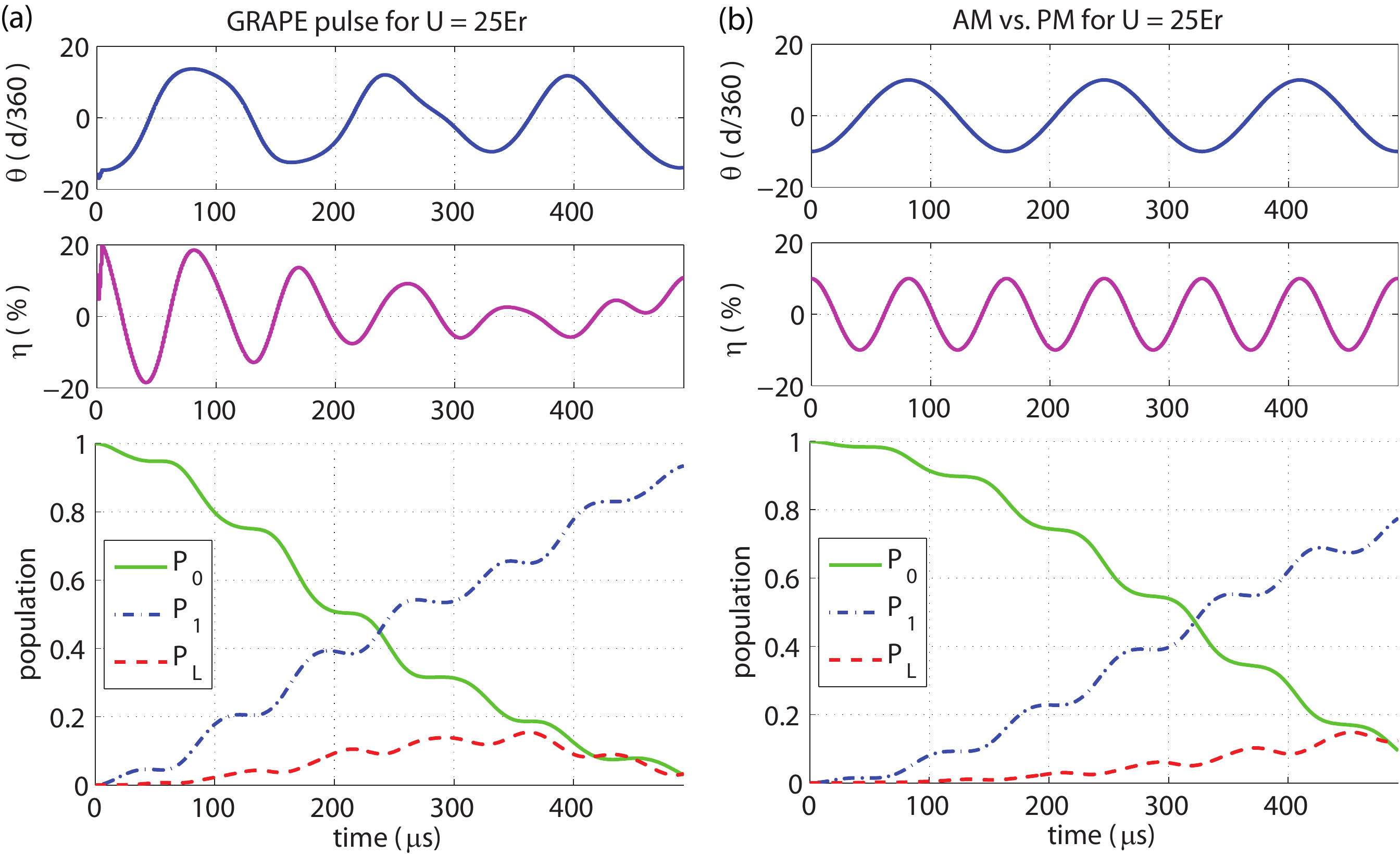}
  \caption{\label{figGRAPEpulse} GRAPE pulse and a similar AM vs. PM pulse. (a) GRAPE pulse and the corresponding simulation results with the lattice Hamiltonian without gravity. (b) A similar AM vs. PM pulse which has the same pulse duration and the variance as the GRAPE pulse and the corresponding simulation results with the lattice Hamiltonian without gravity.}
\end{figure*}

One of the generated GRAPE pulses is experimentally tested here. The top and middle plots in Fig. \ref{figGRAPEpulse}(a) show the GRAPE pulse designed for a lattice with depth of $25E_{r}$.
The bottom plot in Fig. \ref{figGRAPEpulse}(a) shows the simulation results for the designed GRAPE pulse with the Hamiltonian of a $25E_{r}$ lattice without gravity. The simulation results show that the designed GRAPE pulse will give final populations of $P_{0} = 3.0\%$, $P_{1} = 93.7\%$,and $P_{L} = 3.3\%$, if the atom initially occupies the state $|0\rangle$.

It is worth noticing that the designed GRAPE pulse looks similar to the coherent control pulse studied in \cite{TwoPath}. The top and middle plots in Fig. \ref{figGRAPEpulse}(b) show one such AM vs. PM pulse, which has the same pulse duration and center frequency as the designed GRAPE pulse. The main difference between the two is the frequency spectrum of the pulses; in the AM vs. PM pulse, each AM and PM contain only a single frequency component, whereas in the GRAPE pulse, presence of other frequency components in both AM and PM can be seen. Specifically, abrupt shifts are applied at the beginning and at the end of the PM in GRAPE pulse. Also, the relative phase between the AM and PM in the GRAPE pulse differs from the optimal relative phase found in \cite{TwoPath} by $90^{\circ}$. The bottom plot in Fig. \ref{figGRAPEpulse}(b) shows the simulation results for this AM vs. PM pulse, for a $25E_{r}$ lattice without gravity. The simulation results show that the AM vs. PM pulse should give final populations of $P_{0} = 9.5\%$, $P_{1} = 78.0\%$, and $P_{L} = 12.5\%$, if the atom initially occupies the state $|0\rangle$. Comparing these results to that of the designed GRAPE pulse shows that the designed GRAPE pulse outperforms than AM vs. PM in terms of suppressing $P_{L}$ and increasing population transfer into $P_{1}$. On the other hand, the already relatively low $P_{L}$ from the AM vs. PM pulse could mean that the major effect of suppressing $P_{L}$ is due to a mechanism similar to the interference between the two-phonon and the one-phonon transitions.

%\begin{figure}[!ht]
%  \centering
%  \includegraphics[width=\columnwidth]{LatticeDistributionGRAPE.eps}
%  \caption{\label{figLatticeDistributionGRAPE} Lattice depth distribution for the GRAPE pulse experiment.}
%\end{figure}

\begin{figure}[!ht]
  \centering
  \includegraphics[width=\columnwidth]{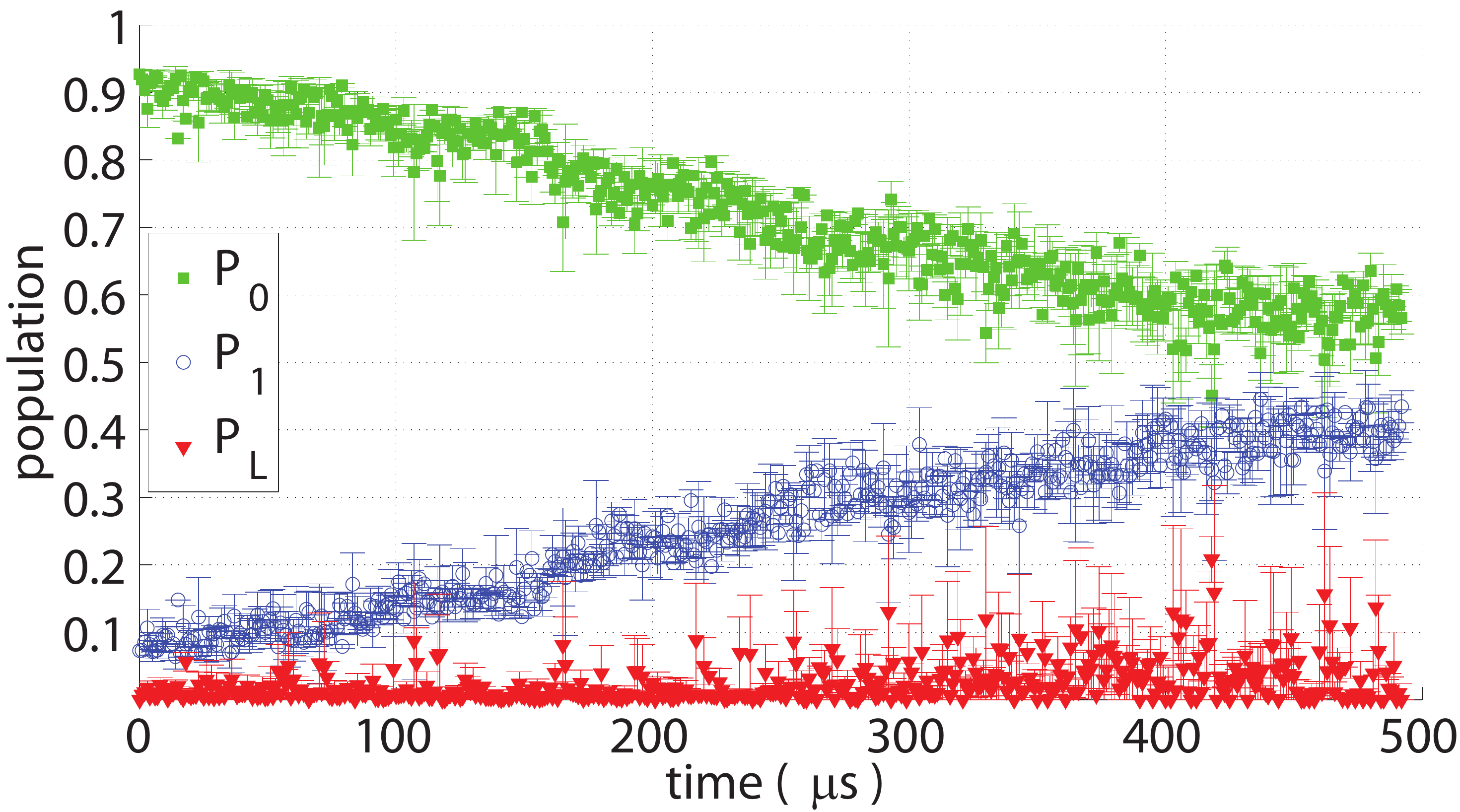}
  \caption{\label{figGRAPEresults} Experimentally measured populations for the GRAPE pulse.}
\end{figure}

Figure \ref{figGRAPEresults} shows the experimentally measured populations as a function of time when the GRAPE pulse in Fig. \ref{figGRAPEpulse}(a) is applied.
The experimentally measured final populations are $P_{0} = 56.6\pm2.4\%$ and $P_{1} = 43.4\pm2.7\%$ with the initial population of $P_{0}^{i} = 94.5\pm1.3\%$ and $P_{1}^{i} = 5.5\pm1.3\%$. The final $P_{L}$ is lower than $3.3\%$ at the $68\%$ confidence level. However, the measured $P_{1}$ is much less than the theoretical predicted value of $93.7\%$. This is due to factors not yet considered in the pulse engineering, such as inhomogeneous broadening and gravity. The pulse was optimized for a lattice depth of $25E_{r}$.
%but due to inhomogeneous broadening, we have a wider range of lattice depths as shown in Fig.\ref{figLatticeDistributionGRAPE}. However, the pulse is sufficiently suppressing $P_{L}$ throughout the duration of the pulse.

\begin{figure*}[!ht]
  \centering
  \includegraphics[width=\textwidth]{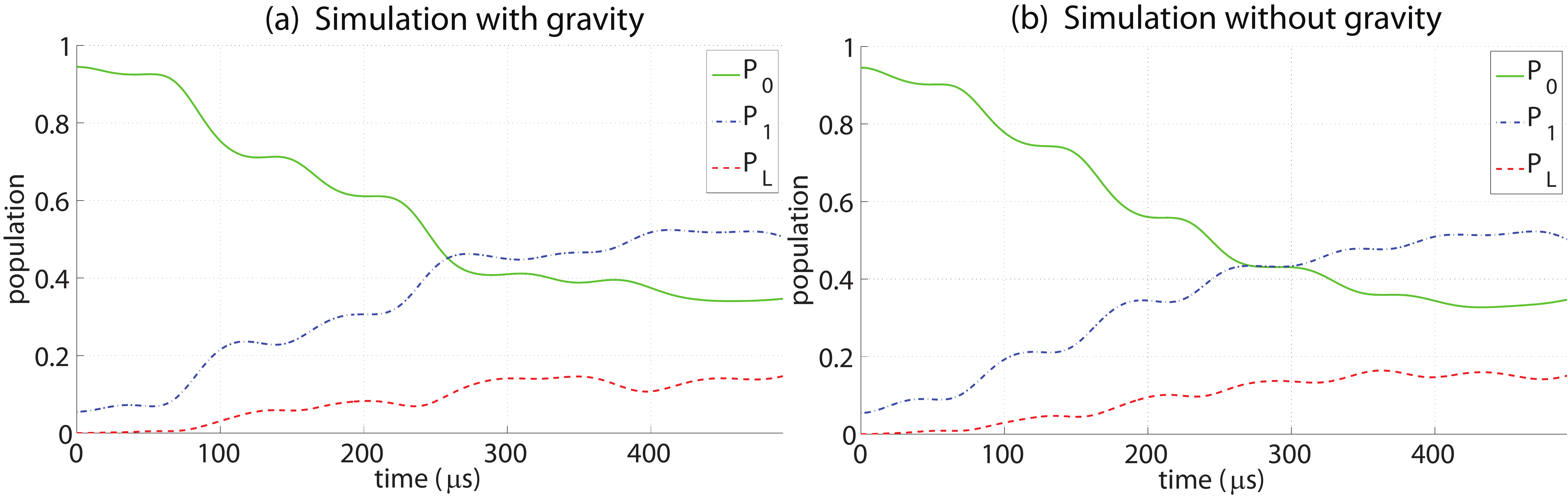}
  \caption{\label{figGRAPESimulation} Simulation results using lattice Hamiltonian for the GRAPE pulse. In (a) the effect of the gravity is considered in the simulation.(b) in the GRAPE pulse result int he absence of gravity. The inhomogeneous broadening is considered in the simulations.}
\end{figure*}

\begin{figure*}[!ht]
  \centering
  \includegraphics[width=\textwidth]{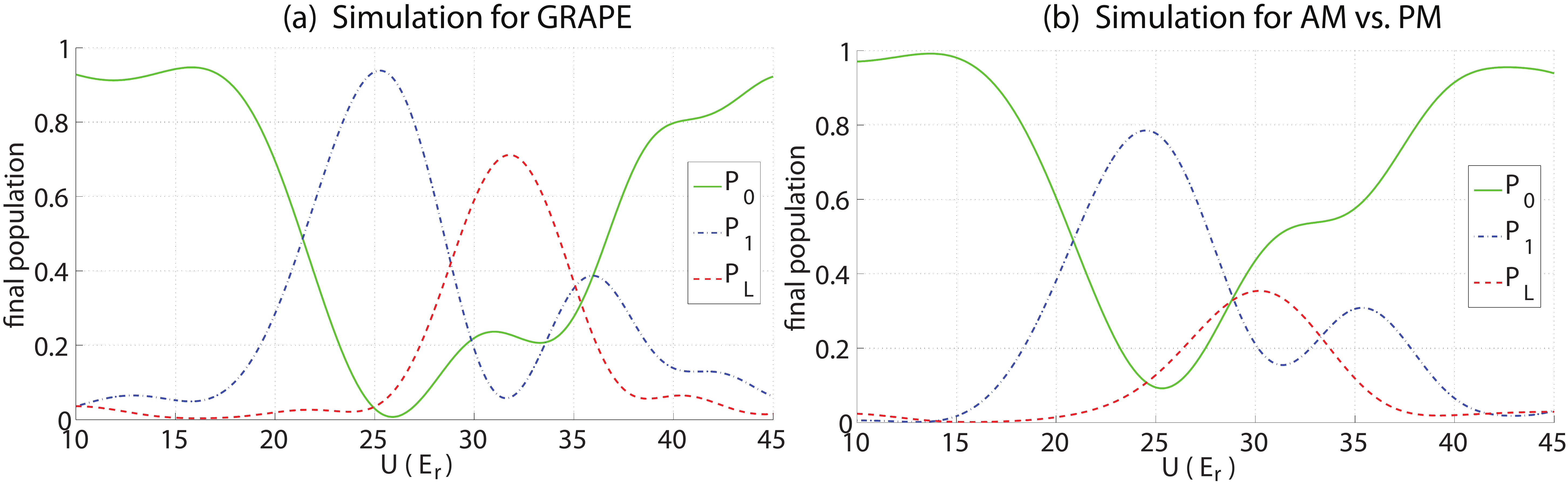}
  \caption{\label{figGRAPEAMPMInhomo} Dependence of the final population on lattice depth. (a) For the GRAPE pulse. (b) For the AM vs. PM pulse.}
\end{figure*}

To further understand the experimentally measured GRAPE results as well as the effects of gravity and inhomogeneous broadening, numerical simulations with the full lattice Hamiltonian are performed taking the initial population distribution into consideration. Fig. \ref{figGRAPESimulation}(a) and (b) show the simulation using the lattice Hamiltonian with and without gravity, respectively, when inhomogeneous broadening is considered.
They both agree relatively well with the experimental results. These results suggest that gravity does not play a major role in the discrepancy. The simulation with gravity shows the final populations to be $P_{0} = 34.7\%$, $P_{1} = 50.6\%$, and $P_{L} = 14.7\%$. The simulation without gravity shows that the final populations to be $P_{0} = 34.7\%$, $P_{1} = 50.2\%$, and $P_{L} = 15.1\%$. We conclude that inhomogeneous broadening is the principle cause of the discrepancy between the measure and expected result of the GRAPE pulse.
%Comparing Fig. \ref{figGRAPESimulation}(a) and Fig. \ref{figGRAPESimulation}(b) shows that the only effect of gravity seems to happen in the duration of the GRAPE pulse.
The performances of the GRAPE and the AM vs. PM pulses in Fig. \ref{figGRAPEpulse} versus different lattice depths are shown in Fig. \ref{figGRAPEAMPMInhomo} (a) and (b), respectively. It can be seen that the GRAPE pulse is more sensitive to deviation of lattice depth from the depth the pulse is optimized for; it greatly suppresses the loss and transfers most of the population to the first excited state at lattice depth of $25E_r$ but this performance dramatically degrades as at the lattice depth of $32E_r$ where around $70\%$ of the population is lost. In contrast, the AM vs. PM pulse never yields a $P_1 > 0.8$ but always keeps the loss to be less $40\%$.

%\begin{figure}[!ht]
%  \centering
%  \includegraphics[width=\columnwidth]{LatticeDistributionGRAPE.eps}
%  \caption{\label{figLatticeDistributionGRAPE} Lattice depth distribution for the GRAPE pulse experiment.}
%\end{figure}

\section{Comparison of different techniques}\label{secComparison}

Before these quantum control techniques could be compared with previously tested techniques, renormalization of the data sets is required. The tests of different quantum control techniques were carried out under slightly different conditions: different distributions of the lattice depths, and more importantly, different initial population distribution between experiments.
%The possibility of testing these techniques all at once is very low due to the stability of the system, e.g., the stability of the lattice laser frequency. The other option is to perform process tomography\cite{} on these different techniques. The possibility of doing process tomography is also low due to the large number of parameters we have tested for each technique. For example, in the complete search part of the one-phonon vs. two-phonon experiment\cite{TwoPath}, we tested 1,000 different parameter sets.
%Because the two conditions mentioned above affect the experimental results, a method needs to be applied to renormalize the different data sets before any comparison could be carried out.
%There are two major factors need to be considered before the different quantum control techniques can be compared: the distribution of the lattice depths and the distribution of the initial population in different states.
The distributions of the lattice depths in different experiments have relatively similar widths and only differ slightly in shape and where they are peaked. Also, the parameters chosen for each technique were optimized for the lattice depth used to test that technique. Hence, the distribution of the lattice depths is not considered when renormalizing the data. The effect of the initial population distributions is relatively easy to correct for. Since the initial population is mostly in $|0\rangle$ with $4\%$ to $8\%$ being in $|1\rangle$, we use a simple model to correct for the effect of this imperfect population initilization: only the change in $P_{1}$ is considered and all populations are renormalized to the initial population in $|0\rangle$. With this simple model, the renormalizing measured populations for different data sets are
\begin{equation}
  \tilde{P}_{0} = \frac{P_{0}}{P_{0}^{i}} \qquad \tilde{P}_{1} = \frac{P_{1}-P_{1}^{i}}{P_{0}^{i}} \qquad \tilde{P}_{L} = \frac{P_{L}}{P_{0}^{i}},
  \label{eqnRenormalize}
\end{equation}
where $\tilde{P}_{0}$, $\tilde{P}_{1}$, and $\tilde{P}_{L}$ are the renormalized population in $|0\rangle$, $|1\rangle$, and leaked states, respectively.

To measure the performance of a quantum control technique, a figure of merit should be chosen. The main goal is to optimize the population transfer from the ground state to the first excited state while minimizing the leakage out of the qubit subspace (the subspace of $P_0$ and $P_1$.) Therefore, the technique with a lower $\tilde{P}_L$ and a higher $\tilde{P}_1$ is superior. In some systems however, the leaked states are very short lived compared to the ground and the first excited state and the population leaked out of the qubit subspace is forever lost. Once such system is the optical lattice used in our experiments, where the lattice is only deep enough to have two bound vibrational states (the qubit subspace) and all higher excited states fall out of the lattice due to gravity in a relatively short time scale. In these systems one wishes to optimize the population inversion $\frac{\tilde{P}_{1}-\tilde{P}_{0}}{\tilde{P}_{1}+\tilde{P}_{0}}$ of the remaining cloud.

\begin{figure*}[!ht]
  \centering
  \includegraphics[width=\textwidth]{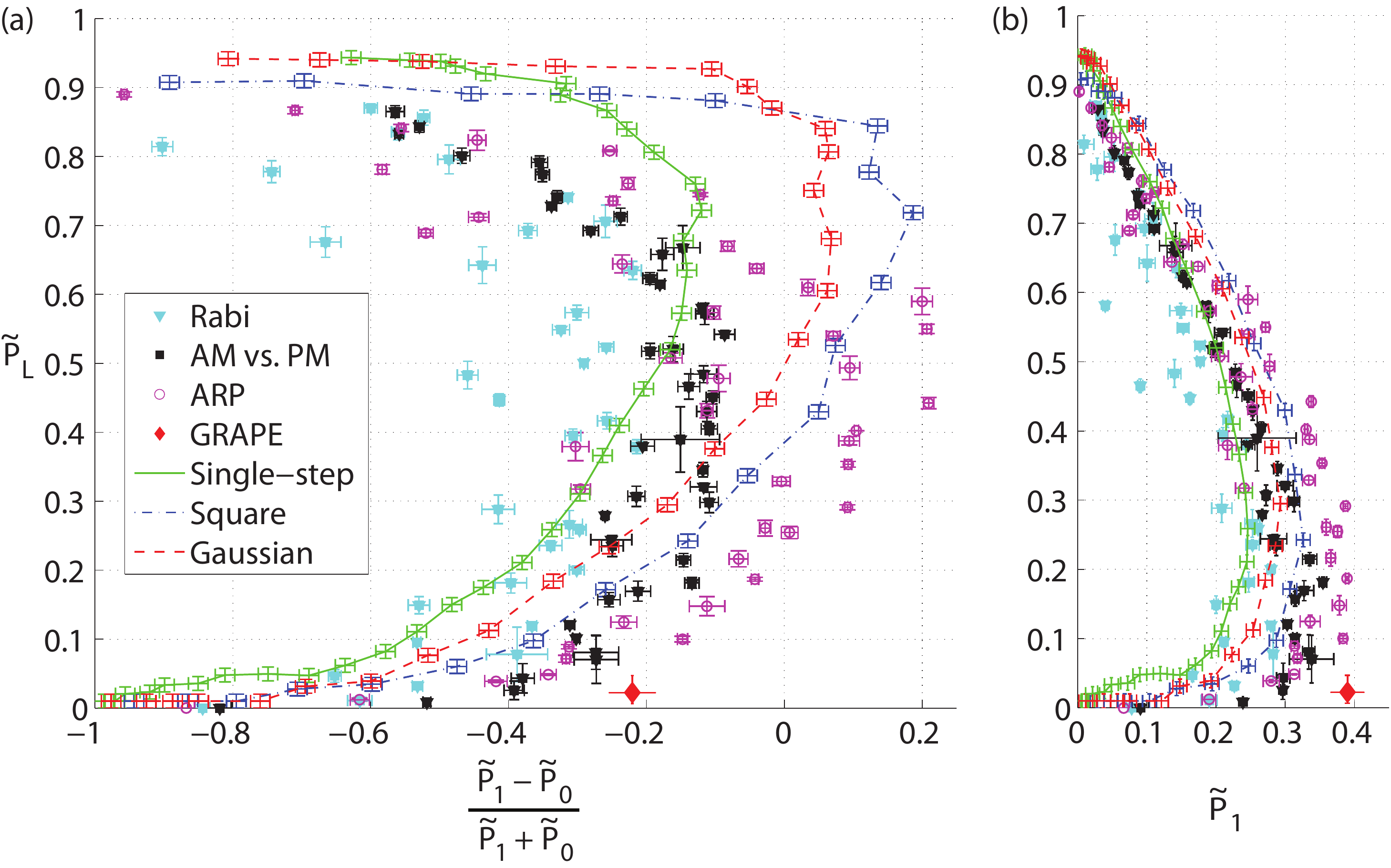}
  \caption{\label{figEverything2Inversion} Comparison of all the tested quantum control techniques. (a) The leakage $\tilde{P}_{L}$ vs. the normalized population inversion $\frac{\tilde{P}_{1}-\tilde{P}_{0}}{\tilde{P}_{1}+\tilde{P}_{0}}$. (b) The leakage $\tilde{P}_{L}$ vs. the increase in the first excited state $\tilde{P}_{1}$.}
\end{figure*}

To compare the different experimentally tested quantum control techniques according to these figures of merit, $\tilde{P}_{L}$ is plotted vs. both $\tilde{P}_{1}$ and the normalized population inversion $\frac{\tilde{P}_{1}-\tilde{P}_{0}}{\tilde{P}_{1}+\tilde{P}_{0}}$ in Fig. \ref{figEverything2Inversion}.
The black solid squares are the results from the AM vs. PM. The magenta open circles are the results from the ARP with positive chirp. The red diamond is the result from the GRAPE pulse. The cyan solid triangles are for a direct monochromatic drive of Rabi oscillations between the two states, which we will not discuss in depth. The green solid line, blue dotted dash line, and red dashed line are results from the Single-step pulse, the Square pulse, and the Gaussian pulse, respectively, taken from \cite{PhysRevA.77.022303}. All the error bars show a $68\%$ confidence interval. Only selected results are plotted to make the figure more readable. 
%The points included are selected as follows: all the results are binned into 0.02 intervals according to the value of $\tilde{P}_{L}$; within each bin, for each technique the point with maximum $\tilde{P}_{1}$ ($\frac{\tilde{P}_{1}-\tilde{P}_{0}}{\tilde{P}_{1}+\tilde{P}_{0}}$) is selected. 
In Fig. \ref{figEverything2Inversion} points with high $\tilde{P}_{1}$ (or high $\frac{\tilde{P}_{1}-\tilde{P}_{0}}{\tilde{P}_{1}+\tilde{P}_{0}}$) and low $\tilde{P}_{L}$ are considered optimal (the points in lower right corner).

It is found that ARP is the best technique for optimizing the normalized population inversion $\frac{\tilde{P}_{1}-\tilde{P}_{0}}{\tilde{P}_{1}+\tilde{P}_{0}}$. As shown in Fig. \ref{figEverything2Inversion}(a), ARP has the highest value of the normalized population inversion: $0.21\pm0.02$. Besides ARP, only the Square and the Gaussian pulses were able to create a population inversion ($\frac{\tilde{P}_{1}-\tilde{P}_{0}}{\tilde{P}_{1}+\tilde{P}_{0}}$) larger than zero. As mentioned before however, only optimizing $\frac{\tilde{P}_{1}-\tilde{P}_{0}}{\tilde{P}_{1}+\tilde{P}_{0}}$ may be insufficient when leakage is important. 

\begin{figure}[!ht]
  \centering
  \includegraphics[width=\columnwidth]{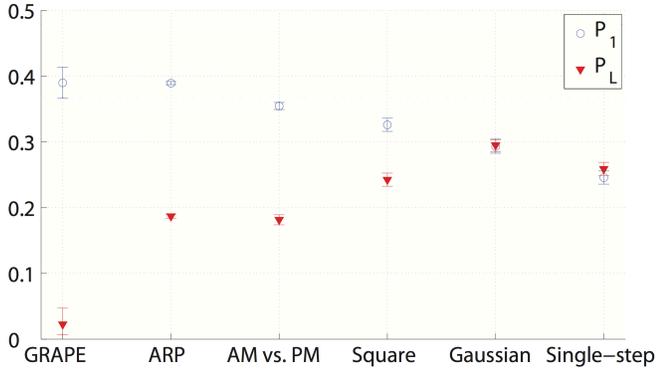}
  \caption{\label{figMaxExt} The highest $\tilde{P}_{1}$ achieved by different quantum control techniques with the corresponding $\tilde{P}_{L}$.}
\end{figure}

When the goal is to increase $\tilde{P}_{1}$ and suppress $\tilde{P}_{L}$ simultaneously, GRAPE is found to be the best among all the techniques we tested. As shown in Fig. \ref{figEverything2Inversion}(b), GRAPE gives the highest $\tilde{P}_{1}$ among all points with the same $\tilde{P}_{L}$ and the lowest $\tilde{P}_{L}$ among all points with the same $\tilde{P}_{1}$. Ranking the techniques in order of highest $\tilde{P}_{1}$ achieved, Fig. \ref{figMaxExt} summarizes the results, indicating both the highest $\tilde{P}_{1}$ for each technique and the corresponding $\tilde{P}_{L}$.
%Alternatively, the performance of different techniques are ranked according to the best $\tilde{P}_{1}$ achieved; the best $\tilde{P}_{1}$ achieved with each technique and the corresponding $\tilde{P}_{L}$ are plotted in Fig. \ref{figMaxExt}.
The AM vs. PM technique achieves almost the same $\tilde{P}_{L}$ as ARP but with slightly lower $\tilde{P}_{1}$. The GRAPE and ARP pulses outperform all the other techniques in terms of $\tilde{P}_{1}$, achieving $38.9\pm0.2\%$ and $39\pm2\%$, respectively. The GRAPE pulse is better in the sense that it transfers almost no population into higher excited states (less than $3.3\%$ of the ground state population, at the $84\%$ confidence level), while $18.7\pm0.3\%$ of the ground state population is transferred into higher excited states by using the ARP.

\section{Conclusion}\label{secConclude}

To conclude, we experimentally tested and compared different quantum control techniques for coupling vibrational states of atoms in an optical lattice. We found that different techniques have advantages when different figures of merit are considered. ARP and GRAPE transfer the most population into the first excited state from the ground state: $38.9\pm0.2\%$ and $39\pm2\%$ respectively. When leakage is taken into account, GRAPE is optimal as it transfers less than $3.3\%$ of the ground state population into the lossy states, while ARP transfers $18.7\pm0.3\%$ of the ground state population into the leaked states. On the other hand, the ARP creates a normalized population inversion (ratio of the difference between the ground state and the first excited state population to the sum of those two) of $0.21\pm0.02$, which is the highest obtained by any of the control techniques we tested; by contrast, GRAPE never produced a population inversion at all.

\begin{acknowledgments}
We acknowledge financial support from NSERC, QuantumWorks, and CIFAR.
\end{acknowledgments}

\bibliographystyle{apsrev4-1}
\bibliography{refs}

\end{document}